# $B_1^+$ mapping near metallic implants using turbo spin echo pulse sequences

Iman Khodarahmi[1], Mary Bruno[1], Ran Schwarzkopf[2], Jan Fritz[1], Mahesh B. Keerthivasan[3]

[1]Department of Radiology, New York University Langone School of Medicine, New York, NY, USA

[2]Department of Orthopedic Surgery, New York University Langone School of Medicine, New York, NY, USA

[3]Siemens Medical Solutions USA Inc., Malvern, PA, USA

Correspondence:

Iman Khodarahmi, MD PhD

Department of Radiology, Musculoskeletal Imaging

New York University Langone School of Medicine

Center for Biomedical Imaging

660 First Ave, Room 223

New York, NY 10016, USA

Email: Iman.Khodarahmi@nyulangone.org

Word count: 4997

# ABSTRACT

**Purpose:** To propose a $B_1^+$ mapping technique for imaging of body parts containing metal hardware, based on magnitude images acquired with turbo spin echo (TSE) pulse sequences.

**Theory and Methods:** To encode the underlying $B_1^+$, multiple (two to four) TSE image sets with various excitation and refocusing flip angles were acquired. To this end, the acquired signal intensities were matched to a database of simulated signals which was generated by solving the Bloch equations taking into account the exact sequence parameters. The retrieved $B_1^+$ values were validated against gradient-recalled and spin echo dual angle methods, as well as a vendor-provided turboFLASH-based mapping sequence, in gel phantoms and human subjects without and with metal implants.

**Results:** In the absence of metal, phantom experiments demonstrated excellent agreement between the proposed technique using three or four flip angle sets and reference dual angle methods. In human subjects without metal implants, the proposed technique with three or four flip angle sets showed excellent correlation with the spin echo dual angle method. In the presence of metal, both phantoms and human subjects revealed a narrow range of $B_1^+$ estimation with the reference techniques, whereas the proposed technique successfully resolved $B_1^+$ near the metal. In select cases, the technique was implemented in conjunction with multispectral metal artifact reduction sequences and successfully applied for $B_1^+$ shimming.

**Conclusion:** The proposed technique enables resolution of $B_1^+$ values in regions near metal hardware, overcoming susceptibility-related and narrow-range limitations of standard mapping techniques.

**Keywords:** $B_1$ mapping, metal implant, turbo spin echo, hardware

# 1. INTRODUCTION

Inhomogeneity of the transmit $B_1^+$ field causes spatial variations in flip angle, leading to image intensity alterations, failure of flip-angle-dependent preparation pulses and errors in quantitative measurements. Spatial variation of the magnetic field produced by the RF coil and the dielectric properties of the body collectively contribute to $B_1^+$ inhomogeneity.

$B_1^+$ mapping is frequently used for a variety of applications including RF shimming in high field and parallel transmit systems (1-3), local specific absorption rate (SAR) assessment (4), design of multi-channel RF pulses (5), and improving the accuracy of relaxation parameters (6-10).

$B_1^+$ mapping techniques can be classified as either magnitude- or phase-based methods. Magnitude-based methods of $B_1^+$ mapping rely on measuring the signal ratio from two flip angles (11-15), stimulated echoes (16,17), two identical RF pulses in the steady state (18), using a 180° signal null pulse (19), or applying preparation pulses (20,21). Phase-based $B_1^+$ mapping methods (22-24), on the other hand, encode the flip angle in the phase of the resulting images. The Bloch-Siegert Shift (BSS)-based methods (25-30) estimate the $B_1^+$ maps by exploiting the change in the frequency, and hence the phase shift, of on-resonance spins subject to off-resonance RF excitation.

$B_1^+$ variations are particularly higher when a metallic object is within the excitation domain (31,32). RF pulses induce eddy currents in the conductive metallic object, which result in $B_1^+$ perturbations (33,34). Despite the need, $B_1^+$ mapping in the presence of metallic implants has not been widely explored yet. Reliable $B_1^+$ mapping near metal requires effective suppression of off-resonance artifacts through the implementation of high transmit-receive bandwidth spin echo-based sequences and ideally multispectral techniques, such as SEMAC (35) and MAVRIC-SL (36).

Gradient echo-based $B_1^+$ mapping techniques inherently fail in the presence of metallic implants. The spin echo (SE) dual angle method (DAM) (11) can mitigate metal-related susceptibility artifacts, however, its long acquisition time, variable sensitivity to small changes in $B_1^+$ and non-uniqueness of the results make it less practical. Spin echo-based BSS $B_1^+$ mapping techniques (25-27) suffer from limited dynamic range when high metal-related off-resonance is present (25).

A previous study (37) implemented the gradient-recalled echo (GRE) DAM (14) formalism to SEMAC acquisitions to calculate $B_1^+$ maps surrounding hip arthroplasty implants, however, the contribution of stimulated echoes in SEMAC signal formation is not accounted for in the DAM and may lead to erroneous estimates.

In this work, we present a new turbo-spin echo (TSE)-based method of flip angle mapping near metallic implants, which exploits signal alterations of TSE acquisitions at various sets of excitation and refocusing flip angles along with apriori knowledge of signal evolution to estimate $B_1^+$. The method along with its SEMAC adaptation was tested in phantom and human subjects without or with metal implants.

## 2. THEORY

### 2.1. $B_1$ estimation

$B_1^+$ field variations affect a TSE pulse sequence by proportionally scaling the excitation and refocusing flip angles (FA). In the presence of a $B_1^+$ scale factor of $B_1$, the signal intensity of a TSE sequence with an excitation FA of $\theta$, and a constant refocusing FA of $\varphi$ can be modeled as: $f(B_1\theta, B_1\varphi, \psi)$, where $B_1 =$ actual FA / nominal FA is the nominal $B_1^+$ scale factor, $f(.)$ is the forward signal model, and $\psi$ represents other imaging and relaxation parameters. For $n$ different sets of excitation-refocusing FA, $B_1$ can be estimated by minimizing the difference between the modeled and acquired signal intensities:

$$\hat{B}_1 = \min_{B_1} \sum_{i=1}^{n} \| f(B_1\theta_i, B_1\varphi_i, \psi) - S(\theta_i, \varphi_i, \psi) \|_2 \qquad [1]$$

with $S(\theta_i, \varphi_i, \psi)$ being the pixel signal obtained by the $i^{th}$ set of excitation and refocusing FA. In practice, the above minimization problem can be solved by creating a database of simulated signal intensities, $\mathbb{D}((\theta, \varphi)_i, B_1)$ with $i = 1 \cdots n$, and maximizing its correlation with the measured signal, $\mathbb{S}((\theta, \varphi)_i)$:

$$\hat{B}_1 = \max_{B_1} \; (\mathbb{S} . \mathbb{D}) \qquad [2]$$

### 2.2. Flip Angle Optimization

The accuracy of $\hat{B}_1$ over a range defined in the set, $\Omega$ depends on the choice of $\theta$ and $\varphi$. To increase the sensitivity of the acquisition to the underlying $B_1$, FA sets were chosen to minimize the variance of $\hat{B}_1$. Similar to the formalism used in Cramer-Rao lower bound (38), this was achieved by minimizing the inversed second derivative of the estimator relative to $B_1$ and is described as the following constrained optimization problem:

$$\min_{\theta, \varphi} \left( \sum_{\mathbb{I}} \frac{\partial^2 (\mathbb{I} . \mathbb{D})}{\partial B_1^2} \right)^{-1} \qquad [3]$$

$$b_1^+{}_{rms}(\theta, \varphi, T_{scan}) \ < \ \epsilon.$$

$\mathbb{I}$ in equation [3] represents a normalized $n$-dimensional signal intensity vector randomly chosen from the set of all possible signal intensities. $b_1^+{}_{rms}$ is used as a subject-independent measure of SAR and is defined as a function of the total RF power ($RF_{total}$) and the total scan time ($T_{scan}$) of the pulse sequence: $b_1^+{}_{rms} = \sqrt{\frac{RF_{total}}{T_{scan}}}$.

## 3.    METHODS

All experiments were conducted on a clinical 3T system (MAGNETOM Vida; Siemens Healthcare GmbH, Erlangen, Germany). Human subjects were scanned after obtaining institutional review board approval and informed consent.

### 3.1.  Signal Simulation and Database Generation

The magnetization evolution of a Carr-Purcell-Meiboom-Gill (CPMG) TSE pulse sequence was modeled for various sets of excitation and refocusing FA by simulating the spin propagation according to the Bloch equations (39). Pulse sequence parameters, including RF pulses and gradient timings and waveforms, and the k-space filling pattern, were imported into MATLAB (The MathWorks Inc., Natick, MA) and C++ Bloch equation solvers.

To avoid the computational burden of full volumetric simulations, one-dimensional simulations along the slice dimension are typically sufficient (8). However, since view angle tilting (VAT) gradients (40), often used in metal imaging to reduce in-plane distortions, are played during readout, we performed a 2D simulation along the slice-selection and readout dimensions.

For a prescribed slice thickness of 3-5 mm, the simulation object was a 2.0-cm thick slab of homogeneous tissue along the slice dimension. The relaxation parameters were matched to the gel for phantom, and to the fat for in-vivo experiments. A total number of 201 points were placed along the slab, resulting in a spatial resolution of 0.1 mm in the slice-selection dimension. The spatial resolution along the readout direction was the readout FOV divided by 20. The temporal resolution of the RF and gradient waveforms used in the simulation was 4 microseconds.

For a given set of $B_1$ and excitation-refocusing FA pairs, each run of the simulation generated the TSE signal evolution along the echo train which could be used to estimate the relative signal intensity at any desired echo time. In practice, since all experimental images were acquired with a proton-density weighting, depending on echo spacing, the k-space center was filled with the third or fourth echo. For comparison purposes, the intensity of the echo train was also simulated by the extended phase graph (EPG) model (41,42). Finally, a database of signal intensity was created by repeating the simulation for a range of $B_1$ values ($\Omega$ = [0.3, 3], steps 0.03) and several excitation-refocusing FA pairs.

## 3.2. Signal Model Validation

The accuracy of the Bloch model was tested by placing a homogeneous gel-filled cylindrical tube (T1 = 1350 ms, T2 = 23 ms) at the isocenter of the magnet with its longitudinal axis along the slice encoding dimension. The isocenter placement of the tube, along with its small diameter relative to the magnet bore ensured $B_1^+$ homogeneity, which was later confirmed using a vendor-provided TurboFLASH-magnetization preparation technique (21). To measure the magnetization evolution, the phase-encoding gradients were switched off, and consequently, no spatial in-plane encoding was applied (43). Sequence parameters included: TSE sequence with 100% VAT in the axial plane, TR/TE = 2s/27ms, echo spacing = 8.88 ms (placing the third echo at the k-space center), voxel size = 0.6 x 0.6 x 5 mm³, BW = 698 Hz/pixel and echo train length (ETL) = 7. Using this setup, the signal evolution curves were compared with those obtained from the Bloch and EPG models.

## 3.3. Flip Angle Optimization and Sensitivity Analysis

Using the formalism of section 2.2, the optimal FA sets were determined by simulating the TSE signal for a wide range of excitation ([30-120°], steps 15°) and refocusing ([60-180°], steps 15°) FA sets using the Bloch model. To achieve the optimal FA sets with the highest accuracy, the $b_{1\,rms}^+$ constraint was satisfied by varying the $T_{scan}$, rather than restricting the total RF power. Since $\mathbb{I}.\mathbb{D}$ in Equation 3 is not easily differentiable, this second derivative was estimated by analyzing the "narrowness" of the $\mathbb{I}.\mathbb{D}$ curve at its peak. Specifically, the $B_1$ estimation error was defined as the width of the $\mathbb{I}.\mathbb{D}$ estimator curve at its peak, measured within a ±5% range.

Using the optimal excitation-refocusing FA sets, Monte-Carlo simulations were performed to assess the sensitivity of the $B_1$ estimation technique to $B_0$ and T2 parameters. A wide range of $B_0$ ([0, 10], steps 0.05 kHz) and T2 ([20, 200], steps 5 ms) were investigated. Assuming a sufficiently high SNR, the Rician noise of the magnitude images was approximated by a Gaussian distribution (44). Realistic Gaussian noise was then added to the simulated signal for each configuration and the mean relative error in $B_1$ estimation was calculated for 1,000 realizations over the range of $B_1$ defined in $\Omega$.

## 3.4. Technique Validation in the Absence of Metal

The accuracy of the proposed technique compared to existing reference $B_1$ mapping techniques was tested using two to four sets of TSE images in four settings:

1. Cylindrical gel: A commercial cylindrical water-based gel (diameter = 12 cm, T1/T2 = 300/50 ms) was imaged using a 20-channel receive-only head coil. Acquisition parameters are summarized in Table 1.

2. Rectangular gel: A rectangular phantom containing $MnCl_2$- and Gadolinium-doped ASTM gel (45) with an electric conductivity of 0.4 S/m was prepared in-house. Relaxation parameters included T1 = 360 ms and T2 = 105 ms, measured using variable flip angle GRE (46) and single-echo-spin-echo techniques, respectively. To generate higher degrees of $B_1$ variation across the field of view, the phantom was placed at the extreme side of the table close to the built-in body transmit coil, where both $B_0$ and $B_1$ fields are expected to be less homogeneous. Furthermore, the gel phantom was imaged with both elliptical (EP) and circular (CP) polarizations of the radiofrequency pulse. Combined 32-channel spine and 18-channel body receive-only coils were used for signal reception.

3. Thigh: The right thigh of a healthy 42-year-old male with no metal in the imaged body part was imaged using an 18/1-channel transmit/receive knee coil.

4. Pelvis: The pelvis of a healthy 42-year-old male with no metal in the imaged body part was imaged using 32-channel spine and 18-channel body receive-only array coils with EP, which is the default polarization in clinical torso imaging.

For all four configurations, axial VAT proton-density weighted images were acquired at the optimal excitation-refocusing FA sets determined in the previous step. The reference $B_1$ mapping techniques with matched acquisition matrices and voxel sizes included:

a. TFL: Vendor-recommended rapid method of $B_1$ mapping based on a TurboFLASH sequence equipped with a preceding RF pulse for magnetization preparation (21) with, FA preparation pulse = 80°, FA TurboFLASH = 8°, TR/TE = 30s/2.13ms (cylindrical gel), 30s/1.9ms (rectangular gel), 10s/2.0ms (thigh), 10s/1.9ms (pelvis), acceleration factor = 2 (GRAPPA), readout BW = 610 Hz/pixel (cylindrical gel), 700 Hz/pixel (rectangular gel, thigh, pelvis), ETL = 176 (cylindrical gel), 128 (rectangular gel), 212 (thigh), 282 (pelvis), total scan duration = 1:00 (cylindrical gel), 1:01 (rectangular gel), 0:20 (thigh, pelvis) min:sec.

b.  GRE DAM: 2D GRE DAM with $\alpha$ and $2\alpha$ ($\alpha = 45°$) excitation FA, TR/TE = 5s/3.03ms (cylindrical gel, thigh), 8s/2.72ms (rectangular gel), 6s/2.72ms (pelvis), acceleration factor = 2 (GRAPPA), readout BW = 500 Hz/pixel, total scan duration = 2 x 8:20 (cylindrical gel), 2 x 10.08 (rectangular gel), 2 x 9:50 (thigh), 2 x 15:18 (pelvis) min:sec. Through-slice dephasing errors of the 2D GRE DAM method were corrected using the method described in (47). Briefly, the Bloch-simulated complex transverse signal was integrated across the slice for a wide range of FA. A $B_1$ lookup table was created of the ratio of the absolute value of the integrated signals at $2\alpha$ and $\alpha$ as a function of the FA (Figure 1A). The ratio of the signal acquired at $2\alpha$ and $\alpha$ at each pixel was compared to the lookup table to obtain the true FA. Subsequently, $B_1$ was calculated as the ratio between the true and nominal flip angles prescribed at the scanner.

c.  SE DAM: 2D SE DAM with CPMG scheme and $\alpha$-$2\alpha$ and $2\alpha$-$4\alpha$ ($\alpha = 60°$) excitation-refocusing FA, TR/TE = 5s/28ms (cylindrical gel, thigh, pelvis), 8s/28ms (rectangular gel), acceleration factor = 2 (GRAPPA), readout BW = 698 Hz/pixel, total scan duration = 2 x 8:25 (cylindrical gel), 2 x 10:16 (rectangular gel), 2 x 9:55 (thigh), 2 x 12:50 (pelvis) min:sec. The through-plane dephasing errors of the 2D SE DAM were corrected similarly (47,48). Here, the lookup table was generated by calculating the ratio of Bloch simulated signal intensity of $\alpha$-$2\alpha$ and $2\alpha$-$4\alpha$ FA spin echo acquisitions (Figure 1A).

The precision of the proposed TSE-based mapping technique was tested in consecutive repeated measurements in the rectangular gel phantom without movement of the MRI table, and repeated measurements in the thigh volunteer after table movement.

## 3.5. Technique Performance and Application in the Presence of Metal

The proposed TSE-based $B_1$ mapping technique was applied in five settings:

1.  Ti-gel: A ceramic-on-polyethylene total hip arthroplasty implant with titanium (Ti) cup and stem placed in the above-described rectangular gel was imaged in axial and coronal planes using VAT (Table 1). For axial VAT, reference mapping techniques were identical to those described above for the rectangular gel. For the coronal plane, matrix- and voxel-matched references included: (a) TFL with FA preparation pulse = 80°, FA TurboFLASH = 8°, TR/TE = 30s/1.9ms, acceleration factor = 2 (GRAPPA), readout BW = 700 Hz/pixel, ETL = 240, total scan duration = 1:00 min:sec. (b) GRE DAM ($\alpha$ and $2\alpha$, $\alpha = 45°$) with through-plane dephasing correction, TR/TE = 8s/2.72ms, acceleration factor = 2 (GRAPPA), readout BW = 500 Hz/pixel, total scan duration = 2 x 17:36 min:sec. (c) SE

DAM ($\alpha$-2$\alpha$ and 2$\alpha$-4$\alpha$, $\alpha$ = 30° and 60°) with through-plane dephasing correction, 60°-120° and 120°-240° excitation-refocusing FA, TR/TE = 8s/28ms, acceleration factor = 2 (GRAPPA), readout BW = 698 Hz/pixel, total scan duration = 2 x 17:44 min:sec.

2. CoCr-gel: A metal-on-metal hip arthroplasty implant with cobalt-chromium (CoCr) cup, head and stem components was embedded in the rectangular gel and scanned in the axial plane using VAT and coronal plane using compressed-sensing (CS) SEMAC (35,49,50) which exploits kz phase encoding to resolve through-plane distortions (Table 1). The parameters for the axial and coronal reference techniques were similar to rectangular gel and coronal Ti gel, respectively.

3. Subject 1: A 33-year-old female with a long Ti femoral intramedullary nail was imaged using axial VAT sequences. The reference standard was TFL with matrix = 256 x 256, voxel size = 0.9 x 0.9 x 4, FA preparation pulse = 80°, FA TurboFLASH = 8°, TR/TE = 30s/1.77ms, readout BW = 610 Hz/pixel, ETL = 1, total scan duration = 0:06 min:sec.

4. Subject 2: An 88-year-old female with a metal-on-metal total hip arthroplasty including CoCr head and Ti cup and stem components was imaged using axial VAT and coronal CS-SEMAC protocols. The reference standard was axial TFL with matrix = 256 x 256, voxel size = 1.2 x 1.2 x 5, FA preparation pulse = 80°, FA TurboFLASH = 8°, TR/TE = 30s/1.82ms, readout BW = 698 Hz/pixel, ETL = 1, total scan duration = 0:30 min:sec.

5. Subject 3: A 53-year-old male with a metal-on-polyethylene total hip arthroplasty including CoCr head and Ti cup and stem components was imaged using VAT sequence in the axial plane. The acquired images were used to estimate the $B_1$ maps of the two transmit channels of a dual transmit system. These $B_1$ maps were combined to perform $B_1$ shimming in the vicinity of the metal hardware, aiming to reduce $B_1$-related artifacts (51).

The mapping techniques were performed using three sets of TSE images in subjects 1 and 2 and four sets in subject 3.

## 3.6. Data analysis

MRI data from all methods were imported into a custom MATLAB toolbox for $B_1$ map generation. To avoid masking any potential method-specific artifact and noise, unfiltered $B_1$ maps were displayed (52). The metal hardware, surrounding areas of off-resonance signal void and the cortical bone in human subjects were manually segmented on TSE images, with masks propagated to other techniques. Expecting a smoothly varying $B_1$ map in non-metal containing gel media and body parts, $B_1$ maps were filtered using a 3 x 3 Gaussian kernel (sigma 0.5) before statistical analysis. Normality of $B_1$ values was tested using Kolmogorov-Smirnov test. Non-parametric statistical metrics including median and interquartile of the relative differences, root mean squared error (RMSE), Kendall's Tau correlation coefficient, and Lin's concordance correlation coefficient (CCC) were used to assess the pixel-wise association between the test method and reference standard. Lin's CCC measures how closely data pairs align with the 45° line through the origin (53).

# 4.  RESULTS

## 4.1.  Signal Simulation and Validation

The echo train signal intensity from experiments, Bloch simulations, and the EPG model for two representative excitation-refocusing FA pairs is shown in Figure 1B. The signal intensity of the Bloch model closely matched the experimental data, outperforming the EPG model. Figure 1C illustrates the normalized signal of a proton-density weighted acquisition across a range of excitation and refocusing FA sets, using experimental data as well as Bloch and EPG models. The Bloch simulator provided more accurate signal intensity estimates, with relative errors of -9.9 to 2.9%, compared to the EPG model which showed relative errors ranging from -10.2 to 17.2%.

The modulation of the signal intensity at the k-space center as a function of $B_1$ (Figure 1D) forms the basis for the proposed $B_1$ mapping technique. Higher excitation-refocusing pairs demonstrate greater signal at low $B_1$ values, whereas lower excitation-refocusing FA pairs generate more signal at high $B_1$ values.

## 4.2.  Flip Angle Optimization and Sensitivity Analysis

Figure 2A illustrates the error in $B_1$ estimates for various permutations of two, three and four excitation-refocusing FA sets (equation 3) sorted in ascending order and color-coded by $b_{1\,rms}^{+}$. For two sets (TSE 2 SETS), the lowest error was 8.8, obtained with optimal excitation and refocusing FA of 30°-60° and 75°-120°. This error decreased to 4.3 with three sets (TSE 3 SETS), achieved with optimal FA sets of 30°-60°, 90°-60°, and 120°-105°. With four FA sets (TSE 4 SETS), the error was further reduced to 3.4, using the optimal FA sets of 30°-60°, 30°-180°, 105°-60° and 120°-105°. In each case, local minima with lower $b_{1\,rms}^{+}$ values are present and can alternatively be used for $B_1$ mapping. Figures 2B and 2C show the relative error of $B_1$ estimation for the optimal FA sets as a function of $B_0$ and T2, respectively. As indicated by the white contour lines in Figure 2B, the relative error remains below 20% for $B_0$ < 1-2 kHz across all two, three and four sets. Regarding T2 dependency, the error in $B_1$ estimation is below 20% at $B_1$ > 0.9 for TSE 2 SETS and at $B_1$ > 0.55 for TSE 3 SETS and TSE 4 SETS.

## 4.3.  Technique Validation in the Absence of Metal

The coefficient of variation (CV) for repeated measurements of the rectangular gel phantom ranged from 3.8% to 5.0%, with excellent Lin's CCC between 0.98 to 0.99. The CV of repeated measurements of the thigh volunteer ranged from 2.2% to 4.6%, again showing high Lin's CCC between 0.90 to 0.92 (Figure S1 and Table S1).

$B_1$ maps measured using TSE 2 SETS, TSE 3 SETS and TSE 4 SETS, along with the reference methods, are shown in Figures 3 and 4, with corresponding histograms of relative $B_1$ differences presented in Figures S2 and S3. Median and interquartile relative differences, RMSE, Kendal's Tau and Lin's CCC between the TSE and reference mapping techniques are reported in Table 2.

In the cylindrical gel (Figures 3A and S2A), the TSE 2 SETS and TFL techniques slightly underestimated $B_1$ compared to other TSE and reference techniques. The median differences between TSE 3 and 4 SETS and reference DAM techniques were ≤ 0.7% (CCC ≥ 0.98). A similar pattern was observed when the rectangular gel phantom was imaged in EP (Figures 3B and S2B), with median differences between TSE 3 and 4 SETS and reference DAM techniques of ≤ 0.6% (CCC ≥ 0.98). The GRE DAM in CP imaging of the rectangular gel phantom (Figures 3C and S2C) underestimated the underlying $B_1$ (median difference = 16.7%, CCC = 0.80-0.81). Apart from this, the median difference between the TSE 3 or 4 SETS and reference SE DAM was low, at 0.3% and 0.2%, respectively (CCC = 0.99 for both).

In vivo, the TSE 3 or 4 SETS demonstrated a high correlation (0.84 to 0.91) with the reference SE DAM. The relative difference between TSE 3 or 4 SETS and the reference SE DAM was 2.0% (CCC = 0.84) and 1.3% (CCC = 0.85) for the thigh (Figures 4A and S3A) and 1.7% (CCC = 0.91) and 1.5% (CCC = 0.91) for the pelvis (Figures 4B and S3B), respectively. However, a poor correlation of 0.35 to 0.58 was observed between TSE 3 or 4 SETS techniques and the GRE DAM.

## 4.4. Technique Performance and Application in the Presence of Metal

Figure 5 shows the $B_1$ maps obtained with various techniques in axial and coronal planes for the Ti-gel phantom using VAT. For the CoCr-gel phantom, axial imaging was performed with VAT, while SEMAC was used in the coronal plane due to the high susceptibility of the CoCr (Figure 6). In both phantoms, the TSE technique revealed high and low $B_1$ estimates around the femoral stem, which were barely detected by TFL. TFL in both phantoms and GRE DAM in CoCr-gel showed noise-degraded regions due to metal artifacts, particularly on coronal images. The SE DAM method with α = 60° showed low values at both medial and lateral aspects of the femoral stem, indicating wrapping in $B_1$ estimation. This wrapping could be corrected using the SE DAM method with α = 30° (Figure 5).

In-vivo $B_1$ maps for subjects 1 and 2 are shown in Figures 7A and 7B, respectively. Smaller panels display magnified maps in close proximity to the metal, where noise-dominated $B_1$ maps produced by TFL are better resolved using the TSE technique. In subject 2, SEMAC was used in the coronal plane, producing maps consistent with the patterns observed in phantoms. The TFL-based map in this case was significantly degraded by noise near the highly magnetic-susceptible prosthesis head and neck.

Figure 8 illustrates the clinical application of the proposed mapping technique in subject 3 through $B_1$ shimming. $B_1$ maps from each of the two transmit channels of a dual transmit system were combined to determine the optimal polarization of the radiofrequency pulse. Subsequent imaging with the optimal polarization showed reduced radiofrequency shading compared to the standard-of-care technique, which uses a constant elliptical polarization for all individuals regardless of the presence of metal implant.

## 5.  DISCUSSION

Current $B_1$ mapping techniques often fail to resolve the widely variable $B_1$ values near metal. This study introduces a new $B_1$ mapping technique through the acquisition of multiple TSE images with varying excitation and refocusing flip angles. The proposed method with three or more FA sets showed excellent agreement with reference techniques in the absence of metal and was successfully implemented with a variety of metal implants. Highly variable $B_1$ values around the metal, incompletely resolved by often time-consuming reference methods, were successfully explored with the proposed technique in a shorter time and are consistent with findings from prior numerical simulations (54). The method was implemented in conjunction with multispectral metal imaging techniques and was successfully used in a human subject to perform $B_1$ shimming in the periprosthetic region.

The TFL technique underestimated $B_1$ relative to the proposed TSE and reference DAM techniques, aligning with prior findings (55). The poor correlation between the TSE technique and GRE DAM in in-vivo cases without metal hardware may stem from type II chemical shift and ghost artifacts on GRE DAM maps. Ghost artifacts, primarily seen on source GRE images acquired at 90° flip angle, originate from incomplete recovery of longitudinal magnetization due to insufficiently long TR relative to in-vivo T1 values (56). The proposed TSE technique demonstrated higher correlations with SE DAM than GRE DAM. Despite through-slice dephasing corrections for both methods, GRE DAM is expected to produce less accurate estimates in the absence of $B_0$ correction (47,57).

The GRE DAM technique, in addition to long acquisition times and susceptibility-related image degradation near metal, suffers from a narrow bijective range. The FA settings for GRE DAM ($\alpha = 45°$) and SE DAM ($\alpha = 60°$) were based on values previously proposed for 3D acquisitions with hard pulses (11,55), but were not optimized for 2D acquisitions aiming at resolving high $B_1$ values. The 2D GRE DAM in Figure 1A holds a one-to-one association with the prescribed FA, $\alpha$, at $\alpha < 110°$. Monte-Carlo simulations determine the exact range of $\alpha$ to maximize the accuracy of $B_1$ estimation (58). However, for simplicity, we may assume $\alpha > 25°$ (corresponding to slope < -0.011 in Figure 1A) provides sufficiently high SNR. These limits along with the prescribed $\alpha$ determine the range of resolvable $B_1$. For example, a nominal $\alpha = 45°$ can resolve $B_1$ between 0.6 and 2.4. The 2D SE DAM technique is relatively immune to metal-related susceptibility artifacts but suffers from an even narrower bijective range. Its one-to-one association with $\alpha$ occurs at 11° to 90°. However, a similar slope of < -0.011 decreases the upper limit to 78°. With these bounds, a nominal $\alpha = 60°$ can resolve $B_1$ between 0.2 and 1.3. Any $B_1$ outside the calculated range will alias into this range, as demonstrated by low values surrounding the femoral shafts seen in Figures 5 and 6. Using a smaller $\alpha = 30°$ for SE DAM extends the range to 0.4 to 2.6, allowing depiction of high $B_1$ values near metal hardware, but at the expense of increased noise (Figure 5). This calculation shows that a two-angle measurement lacks a sufficiently wide bijective range for $B_1$ measurement, supporting our choice of three or more angle measurements.

While accuracy of the proposed technique improves with a higher number of FA sets, this increases the scan time and SAR. As a trade-off between estimation accuracy, SAR, and scan efficiency, three FA sets were chosen as the optimal configuration in all but one of our in-vivo cases. Application of techniques such as compressed-sensing based undersampling and SAR efficient variable refocusing flip angle schedules could allow more FA sets without time penalties. Technique acceleration is particularly beneficial for SEMAC-based $B_1$ mapping acquisitions. In this setting, the technique could also be enhanced by incorporating a dictionary of signals from individual SEMAC partitions for greater accuracy.

All images in this study were acquired with proton-density weighting, placing the k-space center at the third or fourth echo. Using the first echo to fill the k-space center would make the proposed technique independent of relaxation effects, but less sensitive to $B_1$ variations. It is worth noting that the flip angle optimization of Section 3.3 must be revisited if other echoes are used to fill the k-space center. For in-vivo experiments, fat relaxation parameters were used to generate the dictionary of signal intensities. This is justified as most metal implants, hip arthroplasties in particular, are embedded in the fat-containing marrow cavity of the bone. Nonetheless, as shown in Figure 2C, the optimal FA sets make the estimated $B_1$ robust to T2 variations.

$B_1$ estimation errors increase with higher $B_0$ off-resonances (> 2 kHz) (Figure 2B). It can be established from Figure 5 that, at least for the Ti implant, off-resonances exceeding 2 kHz occur only in small regions near the prosthesis head and neck, where $B_1$ variations are generally minimal. In contrast, off-resonance near the femoral shaft, where substantial $B_1$ variations occur due to antenna effects, remains well below 2 kHz. The proposed mapping technique benefits from the spatial separation of high $B_0$ and $B_1$ areas. However, this performance could be further improved by incorporating $B_0$ distributions into the database and matching algorithms.

The varying T1 contributions to image contrast for different FA schemes (59) were incorporated into our Bloch model. However, only a single TR was numerically simulated. This is justified as the model was validated with a TR of 2s, the shortest used in our experiments. For shorter TRs, the model must account for the effects of relaxation parameters on steady-state image contrast. Additionally, while simulations assumed high SNR typical of clinical protocols, the effect of acquisition SNR on estimation accuracy requires further evaluation.

Recently, Iyyakkunnel et. al. proposed a Carr-Purcell spin echo-based method for $B_1$ mapping in the absence of metal (60). This method uses a dictionary-matching approach to estimate $B_1$ from echo signal oscillations in a reduced refocusing flip angle Carr-Purcell echo train. However, high $B_0$ off-resonances significantly reduce the amplitude of the Carl-Purcell sequence, limiting its use in $B_1$ mapping. The current work's use of multiple excitation-refocusing FA sets sensitizes the TSE sequence to larger $B_1$ variations, even in the presence of high off-resonance, making it a better candidate for estimating $B_1$ near metal objects.

Our study had limitations. First, the estimation performance of the proposed technique was compared to three different reference schemes. The lack of a standardized $B_1$ phantom makes it difficult to quantitatively evaluate its robustness. The use of electromagnetic simulations to quantitatively model $B_1$ effects from various metallic objects (61) could serve as a reference standard and is part of the planned future work. Additionally, while the proposed technique models the effect of T2 and T1, it does not account for other confounding factors such as magnetization transfer. The effect of magnetization transfer on estimation performance, especially with higher flip angles, requires further investigation.

## 6.  CONCLUSION

The proposed technique estimates the spatial distribution of the $B_1^+$ field around metallic implants by applying various excitation-refocusing schemes combined with TSE or SEMAC acquisitions for metal artifact reduction. Promising results were obtained, particularly in regions near the metal surface that are undetectable with other mapping techniques.

FUNDING INFORMATION

This research was supported by a Research Seed Grant awarded to I.K. by the International Skeletal Society. I.K. was funded by a Research Scholar Grant from the Radiologic Society of North America (RSCH2018).

## CONFLICT OF INTEREST

# TABLES

Table 1. Pulse sequence parameters for the proposed $B_1$ mapping technique.

| | Cylindrical gel | Rectangular gel, Ti-gel, CoCr-gel | Thigh | Pelvis | Ti-gel | CoCr-gel | Subject 1 | Subject 2 | Subject 2 | Subject 3 |
|---|---|---|---|---|---|---|---|---|---|---|
| Plane Sequence | Axial VAT | Axial VAT | Axial VAT | Axial VAT | Coronal VAT | Coronal SEMAC | Axial VAT | Axial VAT | Coronal SEMAC | Axial VAT |
| FOV [mm] | 220 x 151 | 400 x 200 | 250 x 187 | 350 x 350 | 350 x 328 | 350 x 328 | 240 x 240 | 240 x 240 | 300 x 225 | 250 x 250 |
| Slices Number / Gap [mm] | 1 / 0 | 60 / 2 | 1 / 0 | 1 / 0 | 20 / 2 | 20 / 0 | 3 / 0 | 36 / 1.75 | 26 / 0 | 10 / 16 |
| Matrix size | 256 x 256 | 256 x 256 | 256 x 256 | 256 x 256 | 256 x 256 | 256 x 256 | 384 x 288 | 256 x 205 | 256 x 154 | 256 x 256 |
| Voxel size [mm] | 0.9 x 0.9 x 4 | 1.6 x 1.6 x 4 | 1.0 x 1.0 x 4 | 1.4 x 1.4 x 4 | 1.4 x 1.4 x 4 | 1.4 x 1.4 x 4 | 0.6 x 0.6 x 3 | 0.9 x 0.9 x 5 | 1.2 x 1.2 x 5 | 1.0 x 1.0 x 4 |
| Readout BW [Hz/pixel] | 698 | 698 | 698 | 698 | 698 | 698 | 651 | 501 | 698 | 698 |
| TR [ms] | 5000 | 8000 | 5000 | 6000 | 8000 | 10000 | 2000 | 8310 | 2630 | 3000 |
| TE [ms] | 31 | 28 | 31 | 28 | 28 | 30 | 29 | 24 | 30 | 31 |
| Echo train length | 11 | 11 | 11 | 11 | 11 | 11 | 13 | 6 | 11 | 11 |
| Echo spacing [ms] | 7.84 | 7.10 | 7.84 | 7.10 | 7.10 | 7.62 | 7.16 | 7.84 | 7.60 | 7.84 |
| Acceleration (factor) | GRAPPA (2) | GRAPPA (2) | GRAPPA (2) | GRAPPA (2) | GRAPPA (2) | CS (8) | GRAPPA (2) | GRAPPA (3) | CS (8) | GRAPPA (2) |
| SEMAC partitions | - | - | - | - | - | 20 | - | - | 20 | - |
| VAT (%) | 100 | 100 | 100 | 100 | 100 | 100 | 100 | 100 | 100 | 100 |
| Acquisition duration [min:sec] | 2-4† x 0:55 | 2-4† x 1:04 | 2-4† x 1:00 | 2-4† x 1:30 | 2-4† x 1:44 | 2-4† x 12:50 | 3 x 0:40 | 3 x 2:48 | 3 x 8:05 | 2 x 0:42 |

Abbreviations: Ti, titanium; CoCr, cobalt-chromium; VAT, view angle tilting; SEMAC, slice encoding for metal artifact correction; CS, compressed-sensing.

† The acquisition time was increased by a factor of 2 to 4, for TSE 2 to 4 SETS, respectively.

Table 2. Comparison of the test and reference $B_1$ mapping techniques for phantom and human subjects.

| | RF pulse polarization | Test technique | Reference technique | Relative difference (%) Median (Interquartile) | RMSE | Kendall's Tau correlation coefficient | Lin's concordance correlation coefficient |
|---|---|---|---|---|---|---|---|
| Cylindrical gel | CP | TSE 2 SETS | vs TFL | 1.1 (3.2) | 0.03 | 0.96 | 0.95 |
| | | | vs GRE DAM | -3.7 (2.4) | 0.02 | 0.97 | 0.89 |
| | | | vs SE DAM | -3.4 (2.2) | 0.02 | 0.98 | 0.91 |
| | | TSE 3 SETS | vs TFL | 4.1 (2.6) | 0.03 | 0.96 | 0.90 |
| | | | vs GRE DAM | -0.7 (1.9) | 0.02 | 0.99 | 0.98 |
| | | | vs SE DAM | -0.3 (1.3) | 0.01 | 0.99 | 0.99 |
| | | TSE 4 SETS | vs TFL | 4.1 (2.7) | 0.03 | 0.95 | 0.86 |
| | | | vs GRE DAM | -0.7 (1.8) | 0.02 | 0.99 | 0.98 |
| | | | vs SE DAM | -0.4 (1.2) | 0.01 | 0.99 | 0.99 |
| Rectangular gel | EP | TSE 2 SETS | vs TFL | 2.7 (6.0) | 0.06 | 0.97 | 0.96 |
| | | | vs GRE DAM | 1.0 (5.9) | 0.06 | 0.97 | 0.97 |
| | | | vs SE DAM | 1.6 (5.6) | 0.05 | 0.97 | 0.97 |
| | | TSE 3 SETS | vs TFL | 1.8 (4.8) | 0.05 | 0.98 | 0.98 |
| | | | vs GRE DAM | 0.3 (4.3) | 0.05 | 0.98 | 0.98 |
| | | | vs SE DAM | 0.6 (3.7) | 0.04 | 0.98 | 0.98 |
| | | TSE 4 SETS | vs TFL | 1.8 (4.8) | 0.05 | 0.98 | 0.98 |
| | | | vs GRE DAM | 0.3 (4.3) | 0.05 | 0.98 | 0.98 |
| | | | vs SE DAM | 0.5 (3.7) | 0.04 | 0.98 | 0.98 |
| | CP | TSE 2 SETS | vs TFL | 2.3 (5.6) | 0.06 | 0.97 | 0.97 |
| | | | vs GRE DAM | 16.9 (10.5) | 0.08 | 0.95 | 0.79 |
| | | | vs SE DAM | 0.8 (4.8) | 0.05 | 0.98 | 0.98 |
| | | TSE 3 SETS | vs TFL | 1.9 (4.8) | 0.06 | 0.98 | 0.97 |
| | | | vs GRE DAM | 16.7 (10.3) | 0.07 | 0.96 | 0.81 |
| | | | vs SE DAM | 0.3 (3.5) | 0.04 | 0.99 | 0.99 |
| | | TSE 4 SETS | vs TFL | 1.9 (4.9) | 0.06 | 0.98 | 0.97 |
| | | | vs GRE DAM | 16.7 (10.2) | 0.07 | 0.96 | 0.8 |
| | | | vs SE DAM | 0.2 (3.5) | 0.04 | 0.99 | 0.99 |
| Thigh | CP | TSE 2 SETS | vs TFL | 7.9 (11.1) | 0.07 | 0.77 | 0.57 |
| | | | vs GRE DAM | 5.7 (15.5) | 0.10 | 0.33 | 0.27 |
| | | | vs SE DAM | 5.2 (12.1) | 0.08 | 0.70 | 0.57 |
| | | TSE 3 SETS | vs TFL | 4.6 (4.5) | 0.04 | 0.89 | 0.78 |
| | | | vs GRE DAM | 2.1 (11.2) | 0.08 | 0.44 | 0.35 |
| | | | vs SE DAM | 2.0 (3.7) | 0.04 | 0.87 | 0.84 |
| | | TSE 4 SETS | vs TFL | 3.8 (4.3) | 0.04 | 0.90 | 0.81 |
| | | | vs GRE DAM | 1.3 (11.0) | 0.08 | 0.44 | 0.35 |
| | | | vs SE DAM | 1.3 (3.0) | 0.04 | 0.87 | 0.85 |
| Pelvis | EP | TSE 2 SETS | vs TFL | 10.4 (12.5) | 0.11 | 0.87 | 0.76 |
| | | | vs GRE DAM | 0.7 (20.3) | 0.17 | 0.61 | 0.56 |
| | | | vs SE DAM | 5.3 (10.9) | 0.10 | 0.90 | 0.84 |
| | | TSE 3 SETS | vs TFL | 6.9 (9.0) | 0.08 | 0.93 | 0.86 |
| | | | vs GRE DAM | -1.1 (18.5) | 0.17 | 0.63 | 0.58 |
| | | | vs SE DAM | 1.7 (4.6) | 0.07 | 0.94 | 0.91 |
| | | TSE 4 SETS | vs TFL | 6.7 (9.0) | 0.08 | 0.93 | 0.86 |
| | | | vs GRE DAM | -1.4 (18.0) | 0.17 | 0.62 | 0.57 |
| | | | vs SE DAM | 1.5 (4.4) | 0.07 | 0.94 | 0.91 |

Abbreviations: EP, elliptical polarization; CP, circular polarization; TSE, turbo spin echo; TFL, turboFLASH; GRE DAM, gradient recalled echo double angle method; SE DAM, single echo double angle method; RMSE, root mean squared error.

GRE DAM ($\alpha$ and $2\alpha$) and SE DAM ($\alpha$-$2\alpha$ and $2\alpha$-$4\alpha$) images were acquired with $\alpha = 45°$ and $\alpha = 60°$, respectively.

FIGURES

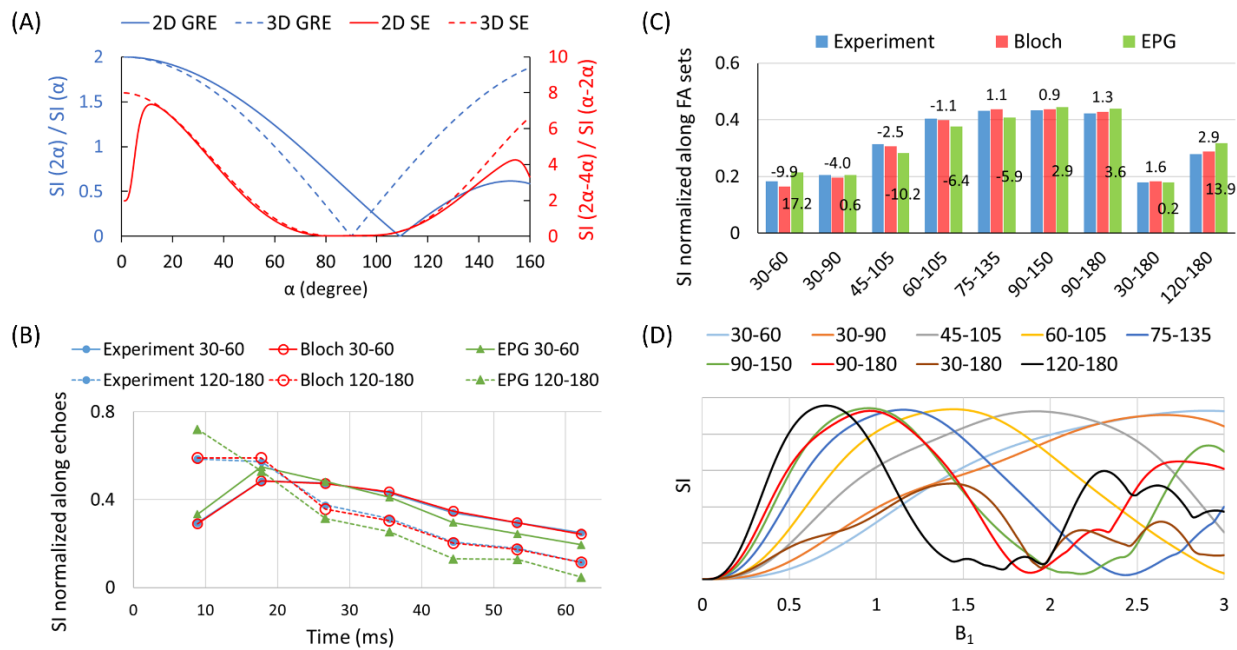

**Figure 1.** (A) Correction of through plane dephasing of 2D gradient recalled echo (GRE) and spin echo (SE) dual-angle B₁ mapping methods. The signal intensity (SI) ratio of two acquisitions with 2α and α (α: excitation flip angle) was simulated by solving the Bloch equations along the slice selection for GRE. Similarly, for SE, two acquisitions with excitation-refocusing flip angles of 2α-4α and α-2α were modeled. Dashed curves represent corresponding 3D ratios with non-selective hard pulses which follow the $2 \cos \alpha$ and $8 \cos^3 \alpha$ dependencies for GRE and SE, respectively. (B) Representative temporal evolution of the echo train signal intensity for two sample excitation-refocusing flip angle pairs (30°-60° and 120°-180°) obtained from experiments, Bloch simulator, and EPG model at B₁ = 1.05 (measure by TFL). The signal intensity was normalized across the echo train. As seen, the Bloch model more closely simulates the experimental data than EPG. (C) Normalized signal intensity of experiments, Bloch simulator, and EPG model at the k-space center (TE = 27 ms, echo # 3) for different excitation and refocusing flip angles. The signal intensity was normalized across the 9 displayed excitation and refocusing flip angle sets. Relative errors (%) of the Bloch and EPG models compared to experimental data are labeled for each flip angle set. (D) Bloch model estimated signal intensity as a function of B₁.

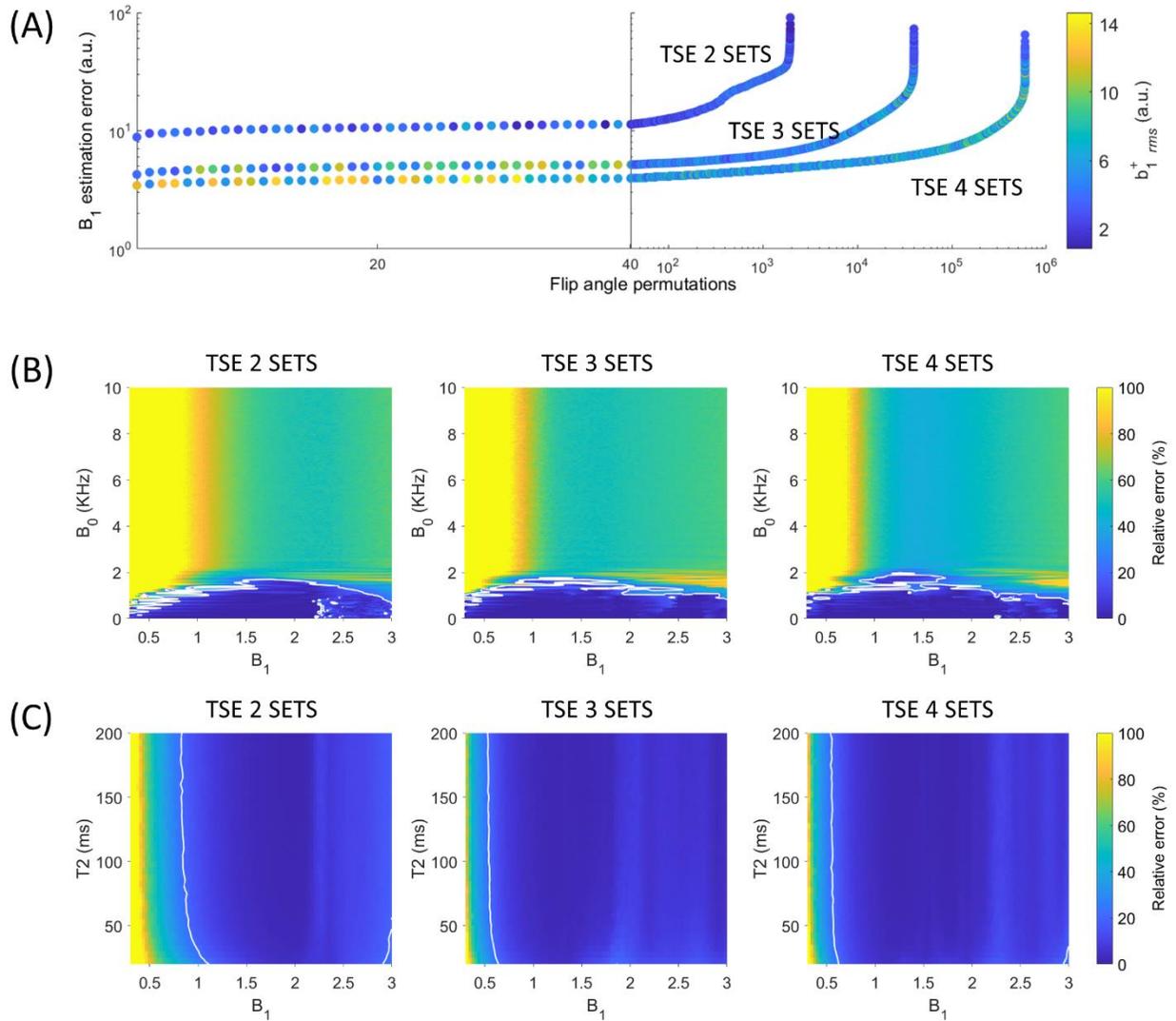

**Figure 2.** (A) Error of $B_1$ estimation for various permutations of two, three and four excitation-refocusing flip angle sets (TSE 2 SETS, TSE 3 SETS and TSE 4 SETS, respectively) sorted in ascending order and color-coded $b_{1\,rms}^+$. (B) Mean relative error in $B_1$ estimation at various $B_1$ and $B_0$ values for optimal two, three and four flip angle sets. White contours demonstrate the 20% relative error. (C) Mean relative error in $B_1$ estimation at various $B_1$ and T2 values for optimal two, three and four flip angle sets. White contours demonstrate the 20% relative error.

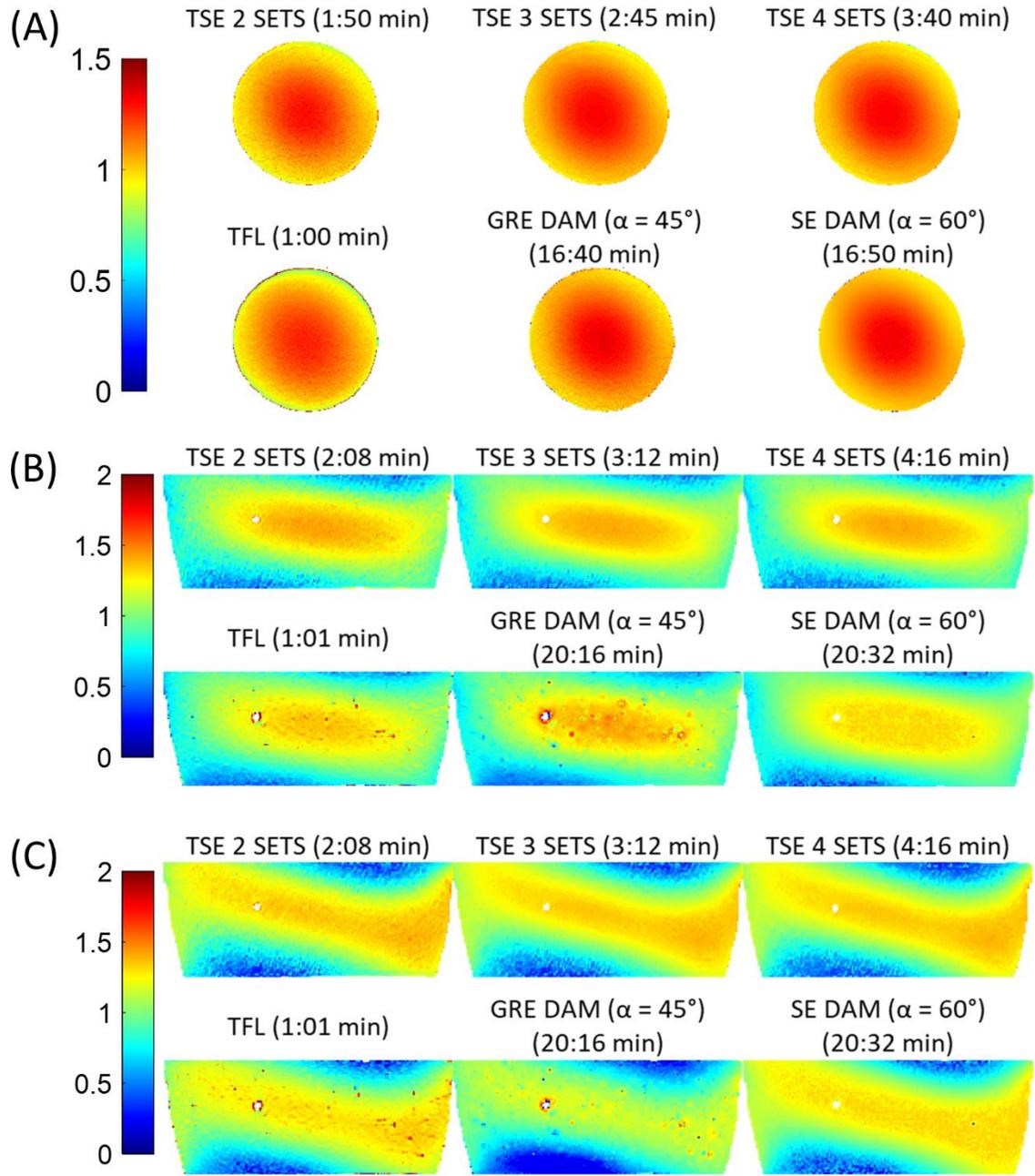

**Figure 3.** $B_1$ maps measured with TSE using two, three and four flip angle sets (referred to as TSE 2 SETS, TSE 3 SETS and TSE 4 SETS, respectively) and the reference TFL, GRE DAM ($\alpha = 45°$) and SE DAM ($\alpha = 60°$) techniques for the cylindrical gel phantom (A), and rectangular gel phantom imaged with elliptical polarization (B) and circular polarization (C) of the radiofrequency pulse.

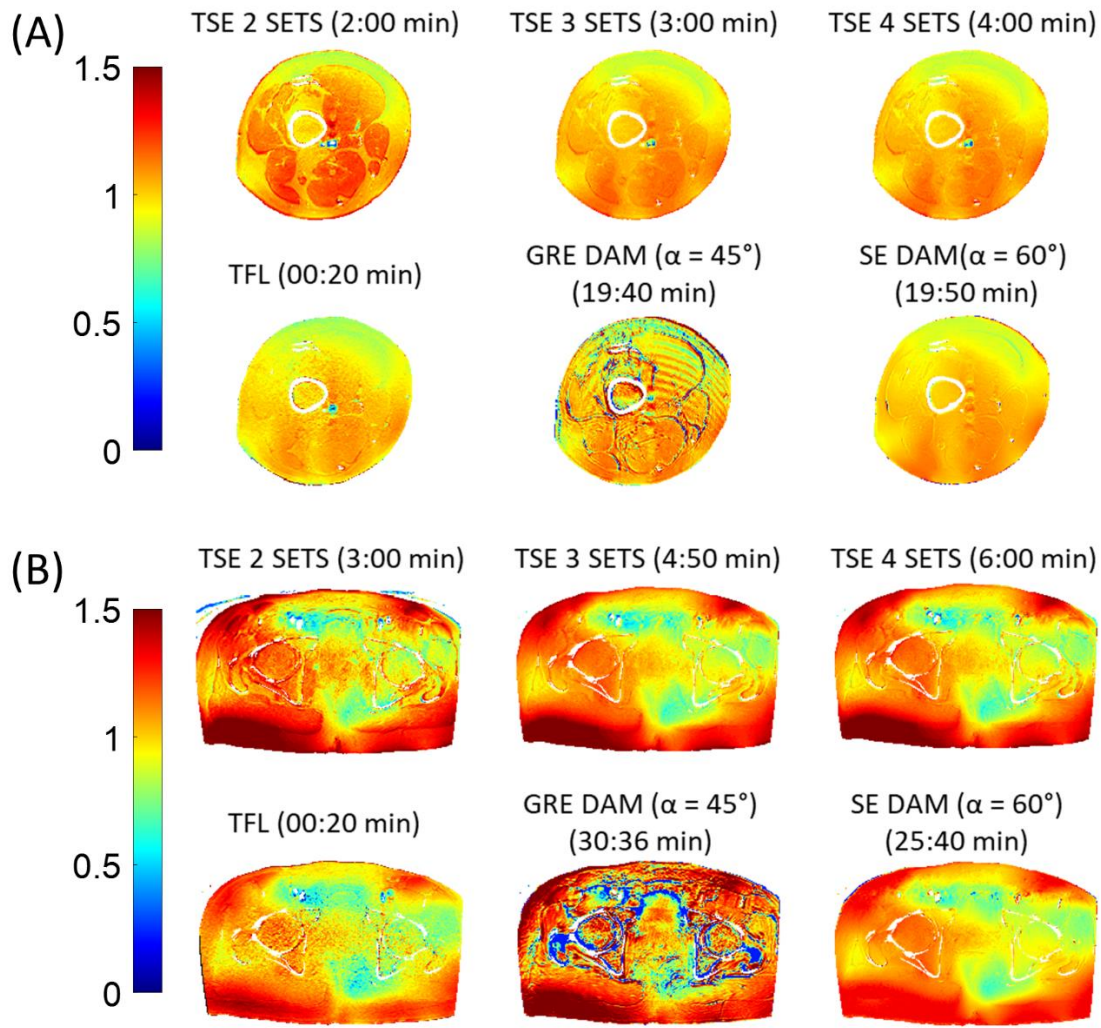

**Figure 4.** B$_1$ maps measured with TSE using two, three and four flip angle sets (referred to as TSE 2 SETS, TSE 3 SETS and TSE 4 SETS, respectively) and the reference TFL, GRE DAM (α = 45°) and SE DAM (α = 60°) techniques for the thigh (A) and pelvis (B) of two separate healthy volunteers.

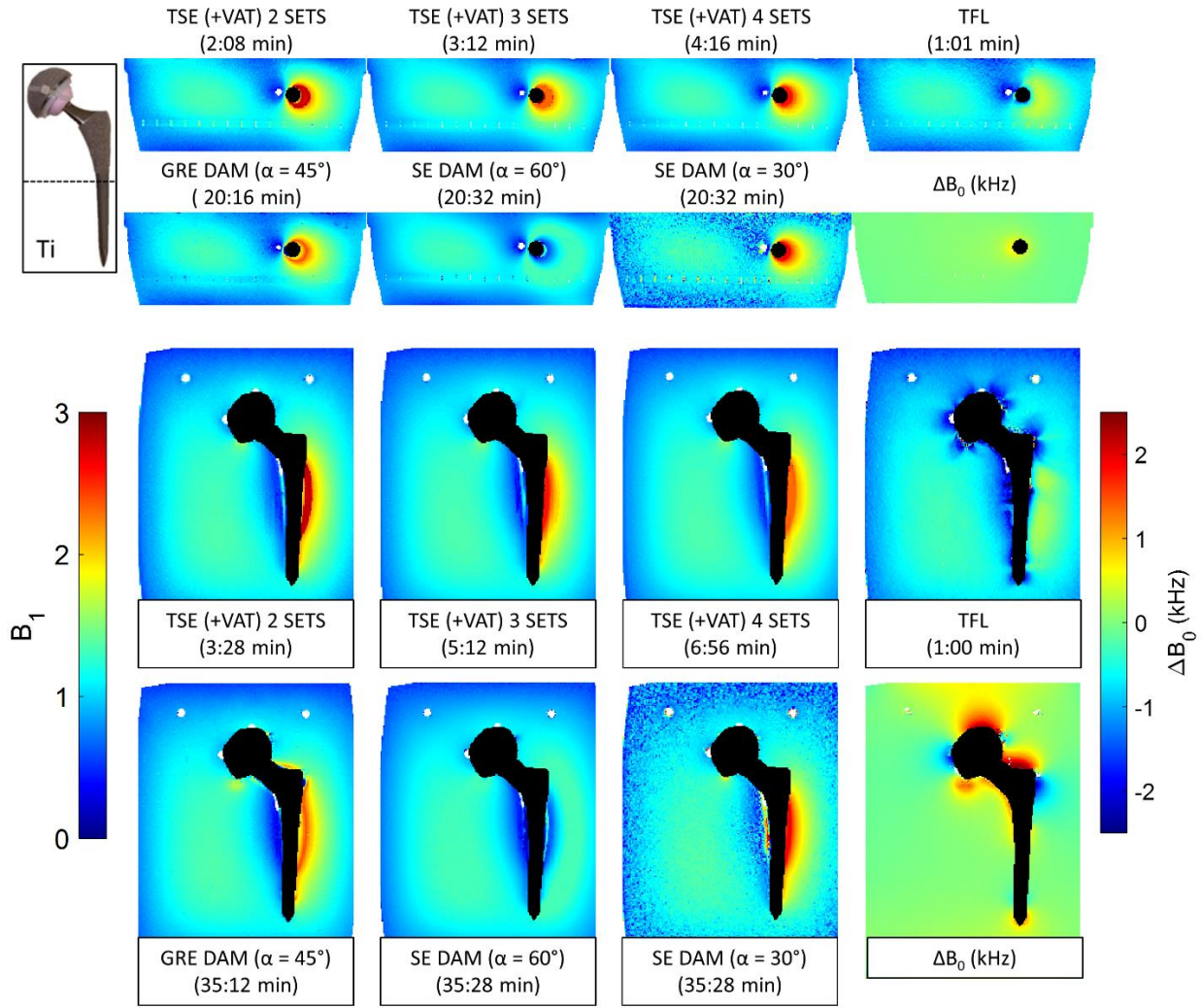

**Figure 5.** $B_1$ and $B_0$ maps of the Ti-gel phantom in axial and coronal planes. Dashed lines on the small panel show the position of the axial planes along the femoral stems. $B_1$ maps are obtained with TSE technique at two, three and four sets using VAT along with the reference mapping TFL, GRE DAM ($\alpha$ = 45°) and SE DAM ($\alpha$ = 30° and 60°) methods.

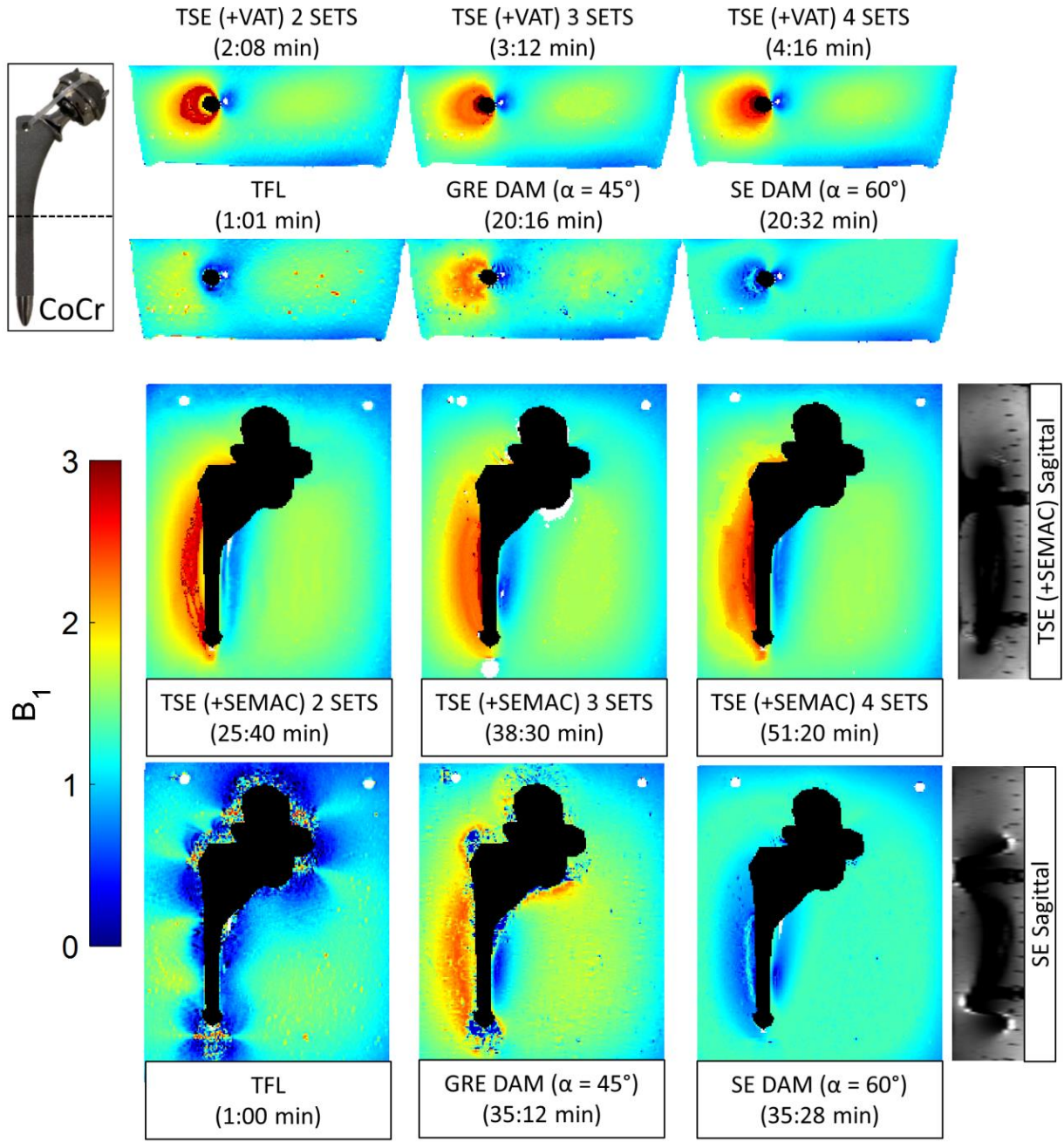

**Figure 6.** $B_1$ maps of the CoCr-gel phantom images in axial plane with TSE VAT and coronal plane with TSE SEMAC at two, three and four sets along with the reference mapping TFL, GRE DAM ($\alpha = 45°$) and SE DAM ($\alpha = 60°$) methods. Dashed lines on the small panel indicate the location of the axial planes along the femoral stems. Grayscale subplots on the left represent SEMAC and SE images reformatted in the sagittal plane which show correction of the SE through-plane geometric distortions with SEMAC.

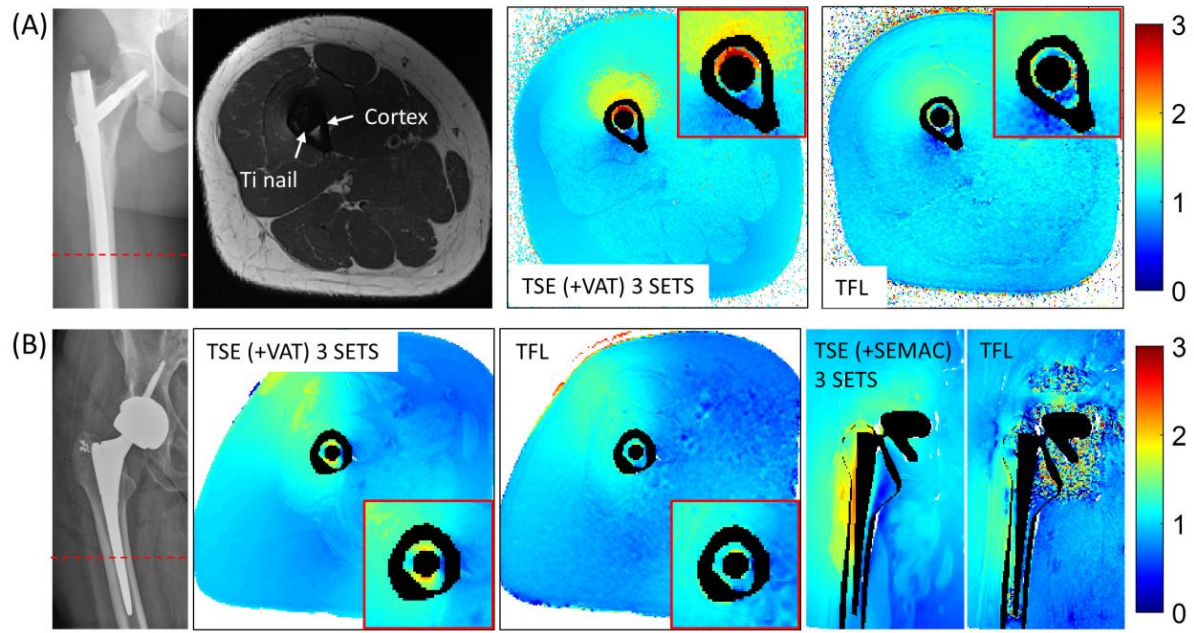

**Figure 7.** (A) Radiographs, axial proton-density weighted VAT image and corresponding TSE (+VAT) and TFL based $B_1$ maps in subject 1 with a titanium femoral shaft nail. (B) Radiographs, axial TSE (+VAT) and TFL based $B_1$ maps, and coronal TSE (+SEMAC) and TFL $B_1$ maps in subject 2 with a metal-on-metal total hip arthroplasty including cobalt-chromium head and titanium cup and stem components. Due to hip flexion only part of the femoral head was included in the coronal image. In both subjects, dashed lines on radiographs indicate the location of $B_1$ maps along the femur. Smaller subplots display magnified $B_1$ maps at the bone-metal interface, highlighting the superior performance of TSE-based methods compared to the noise-dominated TFL maps.

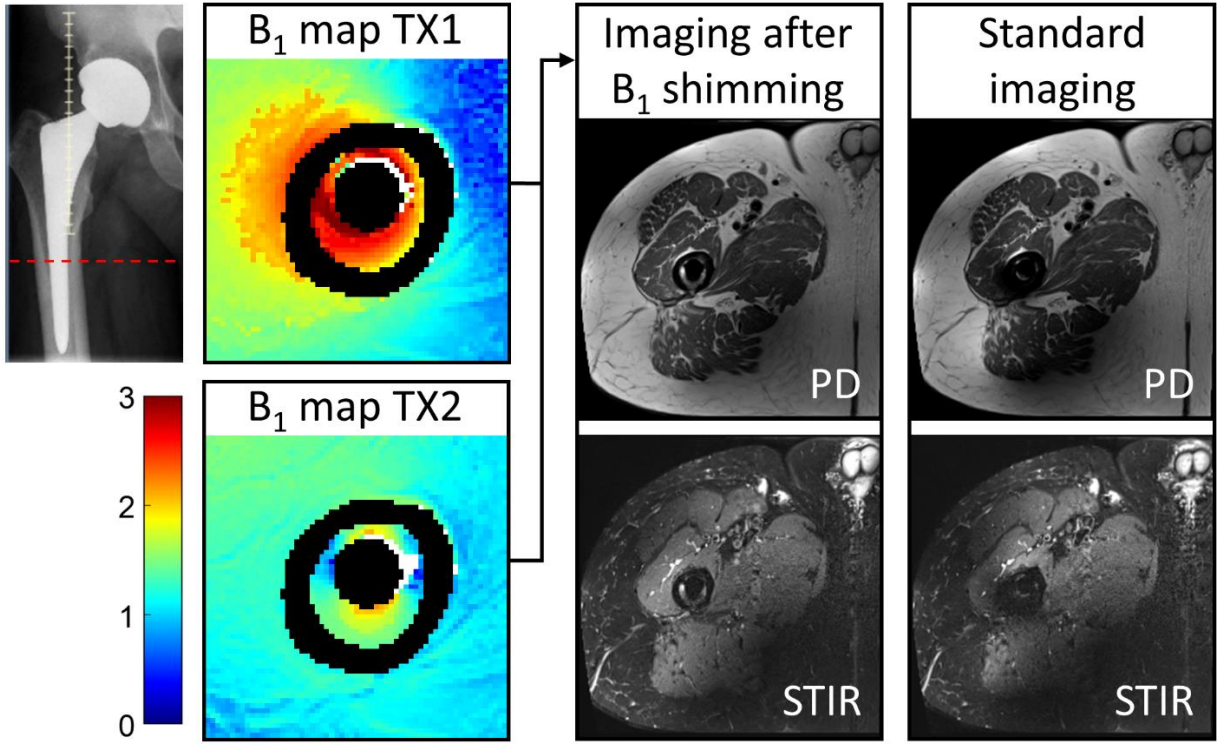

**Figure 8.** Radiographs and $B_1$ maps of each of two transmit (TX) channels of a dual transmit system in subject 3 with a metal-on-polyethylene total hip arthroplasty including cobalt-chromium head and titanium cup and stem components. $B_1$ maps of the bone-metal region were used to determine the optimal polarization of the radiofrequency pulse. Subsequent imaging with the optimal polarization demonstrated reduced radiofrequency shading on both proton-density (PD) weighted and short tau inversion recovery (STIR) images compared to the standard-of-care technique, which applies constant elliptical polarization for all subjects.